\documentclass{article}
\usepackage{graphicx} 
\usepackage{amsmath}
\usepackage{hyperref}
\usepackage{authblk}
\usepackage{tcolorbox}
\usepackage{enumitem}

\title{A theoretical framework for fees in AMMs}

\author{Abe Alexander\thanks{abealexander@outlook.com} \and Lars Fritz\thanks{lsfritz@proton.me}}

\begin{document}

\maketitle
\begin{abstract}
In the ever evolving landscape of decentralized finance automated market makers (AMMs) play a key role: they provide a market place for trading assets in a decentralized manner. For so-called bluechip pairs, arbitrage activity provides a major part of the revenue generation of AMMs but also a major source of loss due to the so-called 'informed orderflow'. Finding ways to minimize those losses while still keeping uninformed trading activity alive is a major problem in the field. In this paper we will investigate the mechanics of said arbitrage and try to understand how AMMs can maximize the revenue creation or in other words minimize the losses. To that end, we model the dynamics of arbitrage activity for a concrete implementation of a pool and study its sensitivity to the choice of fee aiming to maximize the value retention. We manage to map the ensuing dynamics to that of a random walk with a specific reward scheme that provides a convenient starting point for further studies. 
\end{abstract}
\section{Introduction}

Decentralized Finance (DeFi) has emerged as a groundbreaking ecosystem within the cryptocurrency space, offering a wide array of financial services without intermediaries. One of the key strategies employed within DeFi is arbitrage, a practice where traders capitalize on price discrepancies of assets across different platforms or exchanges.

Arbitrage in DeFi operates similarly to traditional finance but with distinct advantages stemming from the decentralized nature of blockchain technology. Participants can exploit price differentials between decentralized exchanges (DEXs), lending protocols, liquidity pools, and other DeFi platforms. These opportunities arise due to variations in supply and demand dynamics, transaction delays, or inefficiencies in pricing algorithms.

One of the primary drivers of arbitrage in DeFi is the concept of composability, wherein various protocols can be seamlessly integrated to create complex financial strategies. For instance, a trader might exploit price disparities between a decentralized exchange and a lending protocol by borrowing assets at a lower rate on one platform and immediately selling them at a higher price on another.

Arbitrageurs play a crucial role in maintaining market efficiency within the DeFi ecosystem. By exploiting pricing inefficiencies, they help align prices across different platforms and contribute to overall market stability. However, arbitrage opportunities are often short-lived as they attract other traders, leading to rapid price convergence.

Despite its potential for profit, arbitrage in DeFi carries certain risks, including impermanent loss, transaction fees, and smart contract vulnerabilities. Moreover, regulatory uncertainties and protocol risks add further complexity to DeFi arbitrage strategies.

In conclusion, arbitrage in DeFi represents a dynamic and lucrative aspect of decentralized finance, allowing traders to capitalize on price differentials across various protocols. While offering significant profit potential, it requires a deep understanding of market dynamics, smart contract mechanics, and risk management principles. As DeFi continues to evolve, arbitrage will remain a fundamental strategy shaping the landscape of decentralized finance.

In the context of automated market makers (AMMs), arbitrage is a major factor in the revenue generation but also loss due to the so-called 'informed orderflow'. Looking over a variety of AMMs reveals that the main source of fee generation are so-called bluechip pairs that exists all over DeFi but also on centralized exchanges (CEXs). It is sometimes assumed that more than $80\%$ of trading activity in those pairs is in arbitrage deals.

For the purpose of this paper, we thus model the trading activity of an AMM based on the assumption that trading is exclusively via arbritrage. To that end we simulate a source of infinite liquidity which provides a quote at every point in time. This quote follows Brownian motion with a drift, while we make not further assumptions. In parallel, we simulate an AMM. We chose for a constant function type of pool but the results can easily be extended to more complex types of AMMs, such as a concentrated liquidity implementation like the Uniswap v3 pools. 

In Sec.~\ref{sec:setup}, we introduce the mathematical framework of an AMM as well as the principles of arbitrage carried out by an agent acting optimally. While there is a general gain from arbitrage, it turns out that through the setting of the fee, the AMM participates in those gains. It turns out, that for a given price mismatch, there is an optimal fee choice which allows the AMM to retain $2/3$ of the gains, while the arbitrageur has to content with $1/3$. Furthermore, we find that the threshold gain in arbitrage processes (threshold defined as the minimal price difference required for successful arbitrage) is proportional to the square of the fees and inversely proportional to the price impact (non-linear price).

We use those insights in Sec.~\ref{sec:randomwalk} to argue that a good effective model to describe things like effective time between arbitrage events and expected rewards is given by a random walk with a specific reward structure. The main finding in this section is that under steady conditions, the actual choice of a fee is relatively unimportant and the gains of the AMM are invariant under this choice. This can be traced back to an interplay between the specific arbitrage reward structure derived in Sec.~\ref{sec:setup} and fundamental properties of one-dimensional random walks.

In Sec.~\ref{sec:simulation} we go to a more microscopic simulation where we do not impose the points of arbitrage but let the arbitrageur find its points in an optimal way. Our main finding is that we confirm the findings of the random walk discussion.

We end with a conclusion and with an outlook, see Sec.~\ref{sec:conclusion}.

\noindent{\it{Related literature:}}

 AMMs can be traced back to \cite{hanson2007logarithmic} and \cite{othman2013practical} with early implementations discussed in \cite{lehar2021decentralized}, \cite{capponi2021adoption}, and \cite{hasbrouck2022need}. Details of implementation are described in \cite{Adams20} and \cite{Adams21} as well as in a very recent textbook \cite{ottina2023automated}.

We study ways to optimize fees based on an arbitrage-only assumption. Uniswap v3 (\cite{Adams21}) addresses this problem by letting liquidity providers choose between different static fee tiers. Other automated market makers have implemented dynamic fees on individual pools, including Trader Joe v2.1 (\cite{mountainfarmer22joe}), Curve v2 (\cite{egorov21curvev2}) and Mooniswap (\cite{bukov20mooniswap}), Algebra (\cite{Volosnikov}), as well as \cite{Nezlobin2023}. Some of the general properties of toxic flow and loss versus rebalancing have been discussed in Refs.~\cite{Faycal1,Faycal2,Faycal3,milionis2024automated,crapis2023optimal,angeris2024multidimensional}

\section{The setup and statement of the problem}\label{sec:setup}

\subsection{The setup}

We consider the interplay between an AMM and a CEX as infinite liquidity source of arbitrage activity.

For simplicity, we assume that the AMM is described as a Uniswap v2 type constant function pool
\begin{eqnarray}
x_{A}*x_B=L^2
\end{eqnarray}
where $x_A$ is the number of tokens of type $A$ and $x_B$ the number of tokens of type $B$. While we choose for a v2-style pool for simplicity, the analysis applies equally to a v3 style pool. 

Returns from swaps are in general subject to price impact. There are, in principle, three prices worth discussing: the spot price, $p_s$, the effective swap price, $p_{\rm{eff}}$, and the spot price after a swap, $p_s'$. The price before a swap, the spot price, is simply defined by
\begin{eqnarray}
p_s=\frac{x_B}{x_A}\;.
\end{eqnarray}
We now consider a swap in which $\Delta A$ tokens are swapped for $\Delta B$ tokens. At this point, the effective price matters according to
\begin{eqnarray}
\Delta B= p_{\rm{eff}} \Delta A= \frac{p_s}{1+p_{iA} \Delta A} \Delta A
\end{eqnarray}
where $p_{iA}$ accounts for the non-linearity of the pricing function, called price impact. In this expression we kept the price impact generic for it to apply to different types of AMMs. In the case of a constant function AMM it is simply given by $p_{iA}=1/x_A$. The swap ends with a new spot price
\begin{eqnarray}
p_s'=\frac{x_B-\Delta B}{x_A+\Delta A}\;.
\end{eqnarray}
There is a reverse swap with roles reversed and a corresponding inverse price impact $p_{iB}$
\begin{eqnarray}
\Delta A= \frac{p^{-1}_s}{1+p_{iB} \Delta B} \Delta B\;.
\end{eqnarray}

\subsection{An arbitrage cycle}

Now we look at one arbitrage cycle. An arbitrage agent observes that token B trades cheaper in the AMM than on the CEX. In concrete terms, this means that $p_s/p_{\rm{CEX}}>1$. 
The reason for arbitrage is obviously free profit. How can the arbitrage agent optimize the profit? To answer this, we need an equation that accounts for all gains and losses in the process. It is important to reiterate that we assume that the CEX is of infinite liquidity.

In the actual process, the arbitrageur takes a flashloan of size $\Delta FL$ (here we assume that it is taken in the denomination token A). For that, he will pay the fee $\Delta FL f_{\rm{fl}}$ with $f_{\rm{fl}}$. On the other Hand, the AMM charges a fee $f$ for swaps, meaning the actual $\delta A$ entering the pool will be 
\begin{eqnarray}
\Delta A=\Delta FL(1-f)\;.
\end{eqnarray}
After the swap to $\Delta B$ is completed, $\Delta B$ is sold at the CEX for 
\begin{eqnarray}
\Delta=\frac{1}{p_{CEX}}\Delta B \;
\end{eqnarray}
at zero price impact.
Overall, this cycle results in the profit
\begin{eqnarray}
P_A&=&\frac{1}{p_{CEX}}\Delta B-\Delta FL(1+f_{fl})-TXN \nonumber \\ &=& \frac{p_s}{p_{CEX}}\frac{1}{1+p_{iA} \Delta FL(1-f)} \Delta FL(1-f)-\Delta FL(1+f_{fl})-TXN
\end{eqnarray}
to the arbitrageur, where $TXN$ accounts for the cost of transactions. In the following, we use the parameter $\alpha=p_s/p_{CEX}$. 

\noindent{\it{Condition for arbitrage:}} We can start by asking when an arbitrage can be performed successfully. Clearly, the question is equivalent to asking when do we have $P_A>0$ as a function of $f$. To answer this, we first rewrite
\begin{eqnarray}
P_A=\Delta FL \left(\alpha \frac{1-f}{1+p_{iA}\Delta FL (1-f)}-(1+f_{fl})-\frac{TXN}{\Delta FL}) \right)\;.
\end{eqnarray}
Arbitrage with $\Delta A>0$ can only happen for $\alpha>1$ (for $\alpha<1$ one can have the opposite arbitrage with $\Delta B$ in). We are assuming for now that transaction costs are negligible ($TXN=0$), meaning we have a condition
\begin{eqnarray}
\alpha \frac{1-f}{1+p_{iA}\Delta FL (1-f)}-(1+f_{fl})>0\;.
\end{eqnarray}
We find that there is a threshold condition on $\alpha$ for a set AMM fee $f$ which needs to be overcome defined by
\begin{eqnarray}\label{eq:cond}
\alpha>\frac{1+f_{fl}}{1-f}\;.
\end{eqnarray}
In the remainder of this paper we will discuss $f_{\rm{fl}}=0$ which is realistic on many platforms.

There is an optimal $\Delta FL^{opt}$ from the point of view of the arbitrageur which allows to maximize the gain $P^A$. This can be determined from 
\begin{eqnarray}
\frac{dP^A}{d \Delta FL}=0\;.
\end{eqnarray}

The solution is given by

\begin{eqnarray}
\Delta FL^{opt}=\frac{1}{p_{iA}(1-f)}\left(\sqrt{\alpha (1-f)}-1\right)
\end{eqnarray}
which is guaranteed to be positive once we fulfil Eq.~\eqref{eq:cond}. Note that this is equivalent to asking which $\Delta FL$ aligns the new spot price after the swap, $p_s'$, with the price of the CEX $p_{CEX}$.
The last thing to check is what is the optimal gain. There we find

\begin{eqnarray}
P_A^{opt}=\frac{\left(\sqrt{\alpha (1-f)}-1 \right)^2}{p_{iA}(1-f)}
\end{eqnarray}
which is positive for situations that fulfil Eq.~\eqref{eq:cond}.
There is another important quantity which is the revenue of the AMM. This is given by
\begin{eqnarray}
R^{opt}=\Delta FL^{opt}f
\end{eqnarray}
which assumes the form
\begin{eqnarray}
R^{opt}=\frac{\sqrt{\alpha(1-f)}-1}{p_{iA}(1-f)}\;.
\end{eqnarray}

\noindent{{\it{Limiting case of small fee $f$:}}}
In general, it is a given that the fee $f \ll 1$ so while this is a limiting case, it is the most relevant case of them all. Furthermore, we will parametrize $alpha=1+\delta \alpha$ where $\delta \alpha$ is the relative price difference $(p_s-p_{CEX})/p_{CEX}$ and generally $\delta \alpha \ll 1$.
In general, this means that we can rephrase Eq.~\eqref{eq:cond} to leading order as
\begin{eqnarray}\label{eq:threshold}
\delta \alpha > f
\end{eqnarray}
which makes intuitive sense. This means that in general, for an arbitrage activity to happen we expect $\delta \alpha \propto f$.
we can use this to approximate the profit of an arbitrageur as well as the revenue of the DEX as 
\begin{eqnarray}\label{eq:rewards}
P_A^{\rm{opt}}&=& \frac{\left(\delta \alpha-f\right)^2}{4p_{iA}} \propto \frac{f^2}{p_{iA}}\nonumber \\ R^{opt}&=&\frac{f\left(\delta \alpha-f\right)}{2p_{iA}}\propto \frac{f^2}{p_{iA}}\;.
\end{eqnarray}

There is another interesting question that can be asked which is whether for a given $\delta \alpha$, there is an optimal choice of fee, $f^{opt}$, that allows maximum value retention. It turns out that the best choice in that case is $f^{opt}=\delta \alpha/2$. For that choice, we find that the AMM retains $2/3$ of the arbitrage gain whereas the arbitrageur has to content with $1/3$. 

\begin{tcolorbox}
{\bf{In summary:}}
\begin{enumerate}
\item{ The threshold for an arbitrage action behaves like $\delta \alpha \propto f$ and inversely proportional to the price impact.}
\item{The potential gain for the arbitrage agent behaves like $P \propto f^2$}
\item{The revenue of the DEX when arbitrage happens scales like $R \propto f^2$}
\item{The maximum part of the arbitrage gains that an AMM can retain is $2/3$ of it.}
\end{enumerate}
\end{tcolorbox}

From this consideration it seems like the best thing for a DEX to do is to just raise the fee $f$ to optimize revenue, but it is not that simple. The fee $f$ determines how often the arbitrage threshold is reached which turns out to be another important quantity. The remainder of the paper is devoted to studying the interplay between these competing effects.

\section{Effective model: A random walk with rewards}\label{sec:randomwalk}

It turns out that the results of the preceding section motivate a very simple but powerful modeling of the the arbitrage activity. The effective model is a version of the well known random walk. This model has countless applications and originated in physics as a microscopic description of Brownian motion and diffusion. However, it also found applications in the field of finance. It is often used to model the Brownian motion of stock prices with a respective volatility and drift. 

In this discussion, we will put a twist on it which allows to understand a heuristic observations from fee dynamics of AMMs: {\it{In calm times, the actual fee setting matters very little to the revenue generation of AMMs}}.

The rules of a random walk are simple. We consider a walker than can go up and down in steps of unit one (it cannot stand still). It does so with probability $[p_{down},p_{up}]$ obeying $p_{down}+p_{up}=1$ meaning it allows for modelling a symmetric random walk as well as versions with drifts. 

\begin{figure}
\begin{center}
\includegraphics[width=0.92\textwidth]{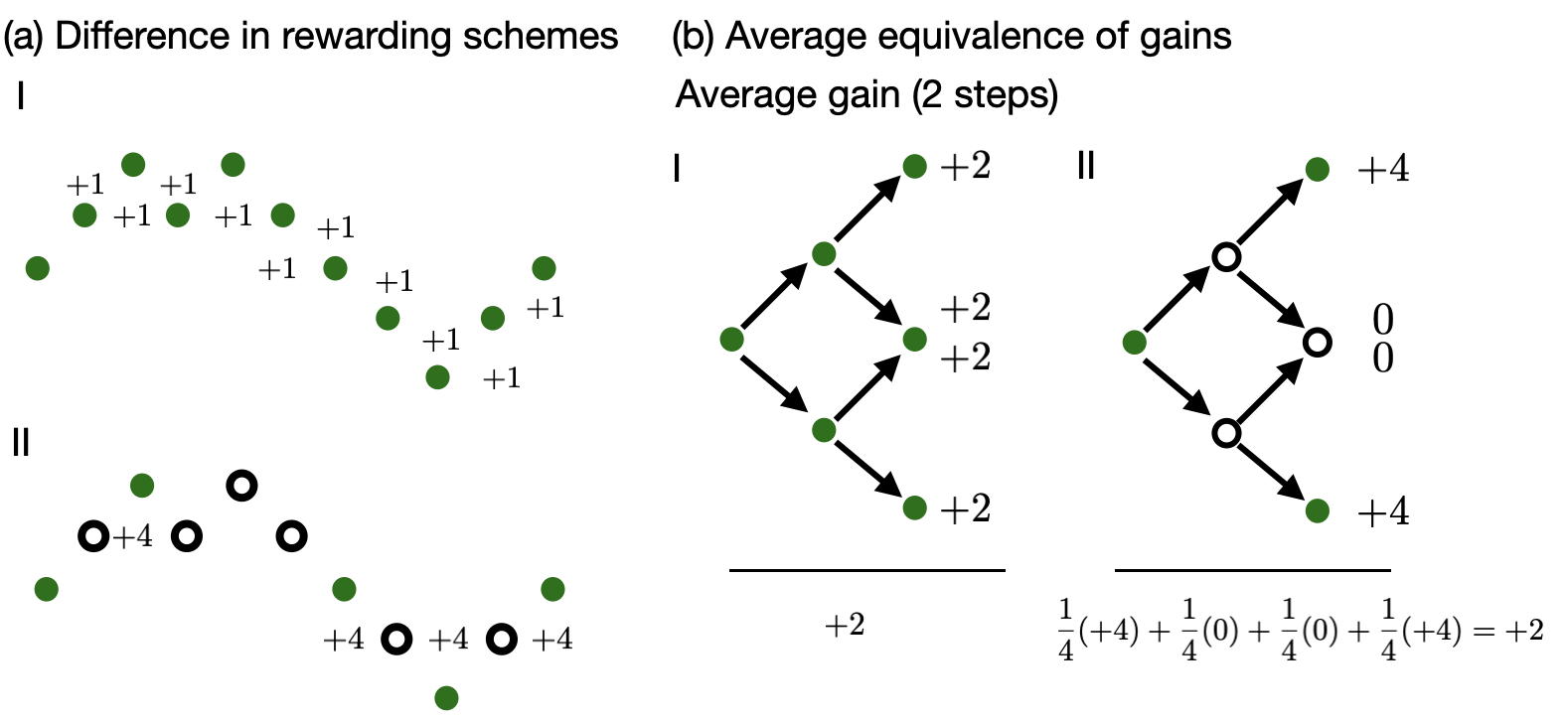}\label{fig:rewardscheme}
\caption{(a) We distinguish two rewarding schemes. (I) rewards every step irrespective of the direction with $+1$ whereas (II) rewards being two levels up or down from the last point of arbitrage with $+4$ rewards. (b) One can easily show that both schemes return the same average gain for sideways motion.}
\end{center}
\end{figure}

We consider a situation that applies in times of low volatility and no preferential drift, so-called 'sideways motion'. The idea is to connect the insights of the previous section with a random walk that can easily be analyzed and simulated. The key insight is this: a step in the random walk corresponds to the price at the CEX either lowering or increasing relative to the AMM. Now there are two situations: a step and the corresponding price difference is big enough to trigger an arbitrage activity. Alternatively, it is not. As we found in Sec.~\ref{sec:setup} we found that the reason can be different fee settings. More concretely, in Eq.~\eqref{eq:threshold} we showed that this arbitrage threshold grows linearly with the fee $f$. It as also shown that while the threshold grows linearly, the arbitrage gains and also the AMM revenue grows quadratically with said fee or threshold, see Eq.~\eqref{eq:rewards}. 

This suggests introducing two different fee schemes, leading to two different rewarding schemes.  Subsequently, we called them strategy I and strategy II  and Fig.~\ref{fig:rewardscheme} (a) makes an attempt at a graphical representation of the key properties.

\begin{enumerate}[label=\Roman*.]
\item{$f_I$ is chosen such that every single step in price is above the threshold, meaning there is an arbitrage activity. We distribute the corresponding reward $+1$.}
\item{$f_{II}=2f_I$ meaning the arbitrage threshold is doubled compared to the previous case. In practice, this implies the last point of arbitrage needs to be two steps higher or lower than the current one to trigger an arbitrage activity. Regarding rewards, this implies that rewards are triggered less frequent, but according to Eq.~\eqref{eq:rewards} the corresponding reward is $+4$}
\end{enumerate}

The question we are now going to answer is this: which strategy is more profitable? Intuitively, the way to the answer is straightforward: If the rewards in strategy (II) are on average triggered faster than every four steps, it wins. If it takes longer, it looses, and if it takes four steps on average, it is identical for practical purposes. 
 
\begin{figure}
\begin{center}
\includegraphics[width=0.8\textwidth]{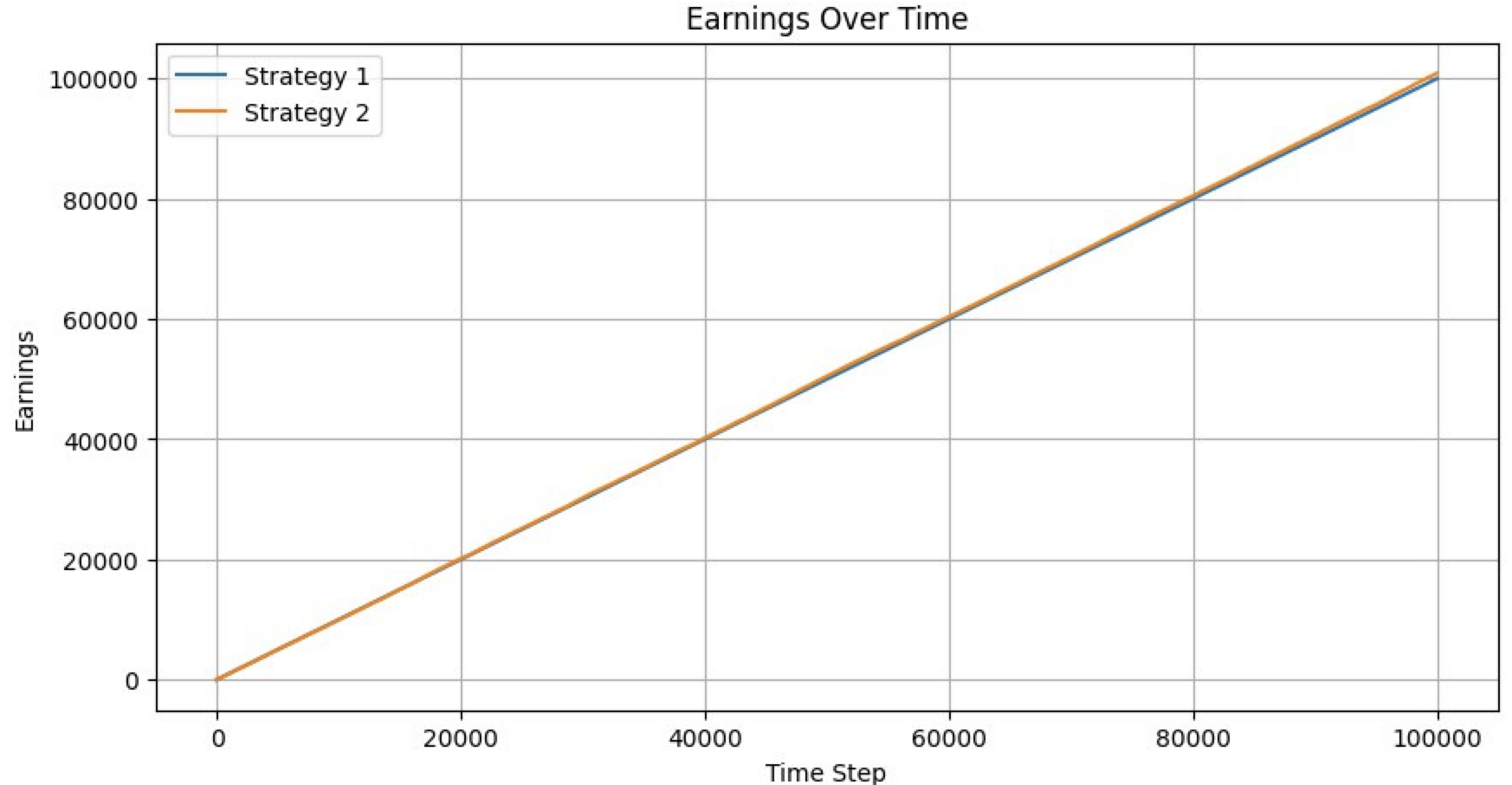}\caption{Random walk with two different strategies giving identical yield}\label{fig:rw}
\end{center}
\end{figure}

We consider the symmetric version ($p_{up}=p_{down}=1/2$) and ran this for $10000$ steps and compared the cumulative return. As expected, the results are identical within errorbars, see Fig.~\ref{fig:rw}.

This can be understood in relative simple terms, see the discussion in Fig.~\ref{fig:rewardscheme} (b): the average gain over two steps is identical in both schemes. 

This discussion hinges on discrete step sizes and reward schemes that are commensurate with the steps. However, the principle at play is more general and can be understood from the random walk again. The mean displacement from the starting point in a random walk is known to scale like

\begin{eqnarray}
\sigma (t) \propto \sqrt{t} \;.
\end{eqnarray}
with time (or number of steps, equivalently).
To stick with our example, this means that if we compare to different fees, $f_I$ and $f_{II}=2f_I$ that define an arbitrage threshold $\sigma_{II}=2\sigma_I$. This implies that the times to reach it are related according to $t_{II}=4t_I$. On the other hand, according to the discussion in Eq.~\eqref{eq:rewards} the fee setting implies that we have rewards $R_{II}=4R_I$. So indeed, time-averaged rewards are identical, as we expected and showed. The advantage of this argument is that it does not require a commensurate rewarding scheme.

\begin{tcolorbox}
{\bf Summary:}

In a situation in which the price moves sideways, the exact setting for fees does not matter (but it can be optimized, as we will show in a follow up study). This is a conspiracy of factors related to the properties of the random walk and the properties of an arbitrage swap which leads to perfect cancellation in the important limit. This statement is borne out by empirical observation. 
\end{tcolorbox}

\section{Comparison against a full numerical simulation}\label{sec:simulation}

\begin{figure}
\begin{center}
\includegraphics[width=1.05\textwidth]{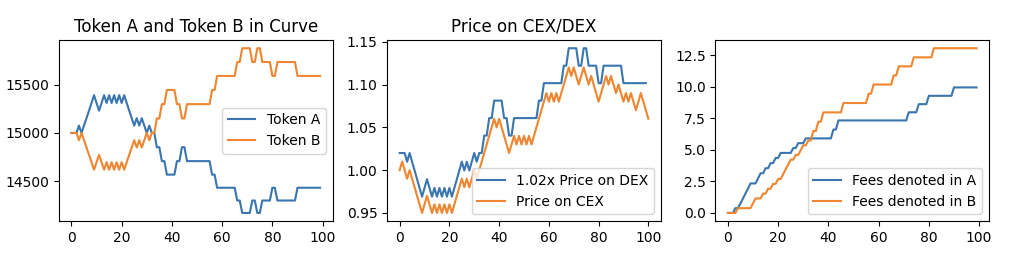}\caption{The simulation was run for $100$ steps with a volatility of $0.01$ per step, an initial price of $p_s=p_{CEX}=1$ and a liquidity of $15000$. The fee setting was $f=0.005$, half the volatility. Left panel: the token of the pool as a function of the price activity; middle panel: A comparison of the price at the CEX and the price at the DEX. Orange curve is the price at CEX, whereas the blue curve is $1.02\times p_s$ to make it distinguishable. We see that the blue curve follows the orange curve as a consequence of the arbitrage activity. Also, we find that certain arbitrage steps are left out because they are not profitable. Right panel: Fees generated in the process.}\label{fig:fullnumerics}
\end{center}
\end{figure}

We simulate an AMM with a pool that is of the Uniswap v2 type and has a relatively high liquidity. The pool is described as a constant product pool of the type
\begin{eqnarray}
x_A x_B=L^2
\end{eqnarray}
and has a starting price $p_s=x_B/x_A$ which is identical to the CEX price $p_CEX$ at this moment in time. Furthermore, there is a fee for swaps that we call $f$. In parallel, we model the price action of the same pair on a CEX. For the purpose of our simulation, the CEX price $p_{CEX}$ will follow Brownian motion. For reference we assume that the starting price of the simulation is $p_s=p_{CEX}=1$ without loss of generality.
The sequence of the simulation is like this: In the following time step, first $p_{CEX}$ will evolve according to a simulated Brownian motion with a specific volatility. Now the code checks whether the misalignment of the prices is sufficient to fulfill 
\begin{eqnarray}
\alpha>\frac{1}{1-f}
\end{eqnarray}
(or the reverse for the other swapping direction), see discussion around Eq.~\eqref{eq:cond}. Now it performs the optimal swap and aligns the price of the AMM with $p_CEX$. Afterwards, the steps repeat.
Throughout, we assume the CEX has infinite liquidity and therefore any buys and sells have no price impact. 
This is the whole setup of the numerical experiment and we can now run it with a variety of parameter settings. What we are tracking are the fees that are generated by the DEX as a function of time and we want to explicitly study the connection between the fee setting and the volatility. There is the understanding that larger volatility requires larger fees in the community. We will show that while this is true in the short term limit, in the long term limit the differences level out and the average revenue generation through fees does not depend on the actual fee setting, in line with the discussion utilizing the simplified random walk.
Before we start with the actual discussion of the statistics we want to show the data as collected in one run for a specific example. The results are shown in Figure.~\ref{fig:fullnumerics}.

In the following we are going to study in which region the conditions defined for the random walk apply. Remember that the main result: {\it{For sideways motion, the exact fee setting did not seem to matter}}. We will further substantiate this claim and point out when it actually holds based on a more 'microscopic' analysis. To this end, we have studied the following case: we simulate an equidistant price series of $1000$ points with a starting price of $p_s=p_{CEX}=1$ and a volatility per time step of $\sigma=0.001$. The liquidity of the system is chosen to be $L=15000$. We have checked that those starting parameters are not important to the statements we are going to make. In a next step we have simulated $1000$ statistically independent runs and studied the role of the fee. We find that the average return for a given volatility $\sigma$ as a function of the fee $f$
 \begin{figure}\label{fig:returnfromfee}
\begin{center}
\includegraphics[width=1.05\textwidth]{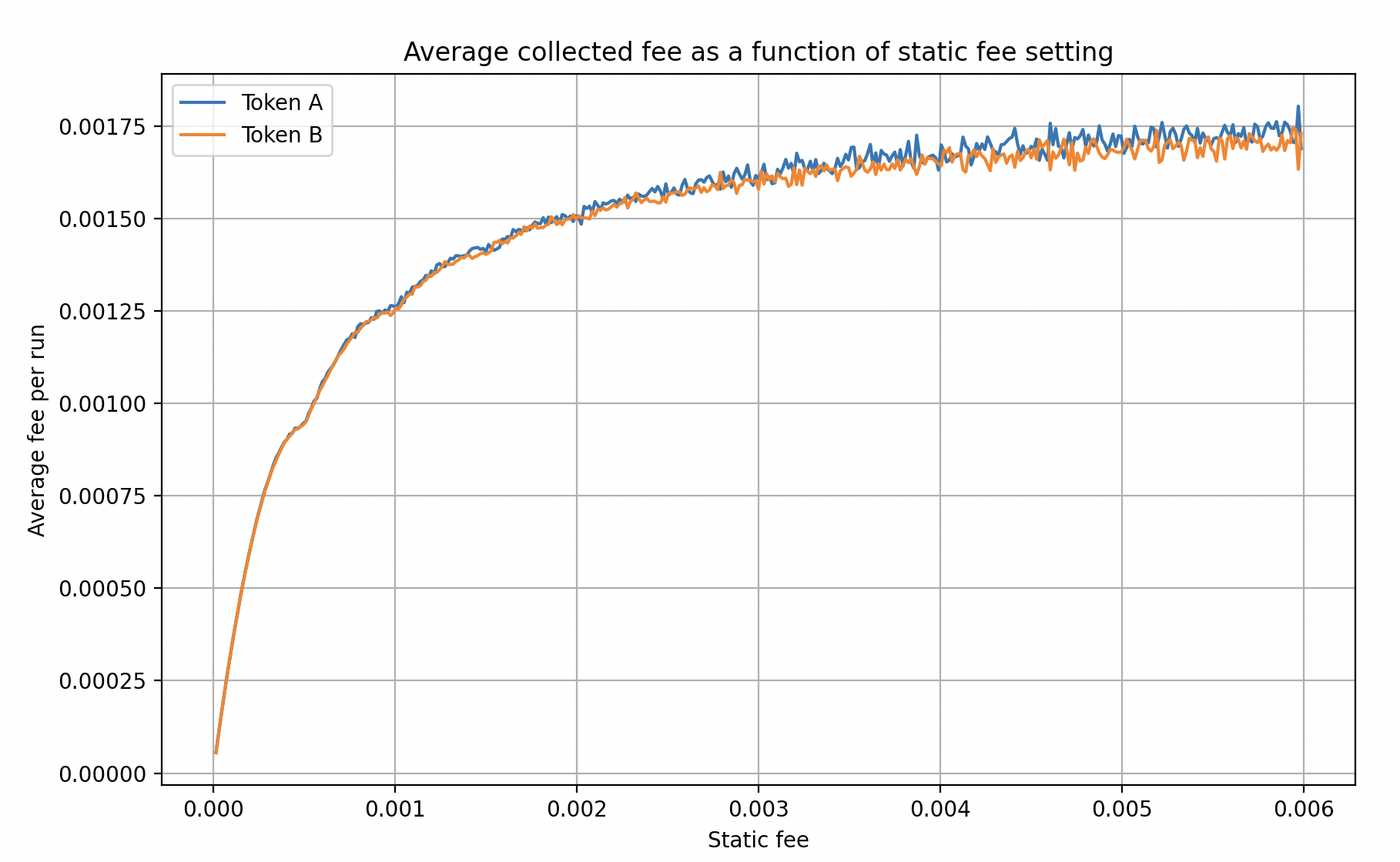}\caption{The simulation was run for $1000$ steps with a volatility of $0.001$ per step, an initial price of $p_s=p_{CEX}=1$ and a liquidity of $15000$. We find that it is a reasonable choice to choose the fee according to the volatility, ideally slightly higher. }
\end{center}
\end{figure} 
is quite sensitive to fees for $f<\sigma$ while it levels off for fees higher than the volatility, see Fig.~\ref{fig:returnfromfee}. The important point is that this validates our previous analysis while giving an indication for a lower bound for the optimal choice of the fees.

\section{Conclusion}\label{sec:conclusion}

In this paper we developed a minimal model to model the return of a DEX for a core pool of heavily arbitraged pairs. We find that for 'sideways motion' the specific value of the fee is not important, as long as is is not chosen to be too high. We also find that a lower bound for the fee is the minimal price change between blocks. We confirm our analysis with both a full fledged numerical simulation as well as a minimal model based on a random walk with rewards. In a subsequent study we will study the influence of asymmetric fees in the case of directed price action.

{\small
  \bibliographystyle{ACM-Reference-Format}
  \bibliography{references}
}

\end{document}